\documentclass[journal]{IEEEtran}
\usepackage{cite}
\usepackage{amsmath,amssymb,amsfonts}
\usepackage{algorithmic}
\usepackage{graphicx}
\usepackage{textcomp}
\def\BibTeX{{\rm B\kern-.05em{\sc i\kern-.025em b}\kern-.08em
    T\kern-.1667em\lower.7ex\hbox{E}\kern-.125emX}}
\usepackage{booktabs}
\usepackage{xcolor}
\usepackage{epsfig}
\usepackage{epstopdf}
\usepackage{acronym}
\usepackage{tikz}
\usepackage{balance}

\usetikzlibrary{shapes,arrows}
\usetikzlibrary{positioning,calc}

\graphicspath{{fig/}}

\def\H{^{\rm H}}

\newacro{SISO}{soft-input soft-output}
\newacro{CP}{cyclic prefix}
\newacro{OCDM}{Orthogonal Chirp Division Multiplexing}
\newacro{OFDM}{Orthogonal Frequency Division Multiplexing}
\newacro{ISI}{inter-symbol interference}
\newacro{MMSE-PIC}{minimum mean squared error with parallel interference cancellation}
\newacro{CWCU}{component-wise conditionally unbiased}
\newacro{MSE}{mean squared error}
\newacro{LLR}{log-likelihood ratio}
\newacro{QAM}{quadrature amplitude modulation}
\newacro{FFT}{fast Fourier transform}
\newacro{LTE}{long-term evolution}
\newacro{FER}{frame error rate}
\newacro{CRC}{cyclic redundancy check}
\newacro{CSI}{channel state information}
\newacro{PFE}{perfect feedback equalizer}
\newacro{GFDM}{Generalized Frequency Division Multiplexing}
\newacro{MIMO}{multiple-input multiple-output}
\newacro{FSC}{frequency-selective channel}
\newacro{FFT}{fast Fourier transform}
\newacro{SISO}{soft-input soft-output}
\newacro{FER}{frame error rate}
\newacro{ETU}{extended typical urban}

\usepackage{soul,color}

\usepackage{tikz}
\usetikzlibrary{shapes,arrows}
\usetikzlibrary{positioning,calc}

\tikzset{add/.style n args={4}{
		minimum width=6mm,
		path picture={
			\draw[black] 
			(path picture bounding box.south east) -- (path picture bounding box.north west)
			(path picture bounding box.south west) -- (path picture bounding box.north east);
			\node at ($(path picture bounding box.south)+(0,0.13)$)     {\tiny #1};
			\node at ($(path picture bounding box.west)+(0.13,0)$)      {\tiny #2};
			\node at ($(path picture bounding box.north)+(0,-0.13)$)        {\tiny #3};
			\node at ($(path picture bounding box.east)+(-0.13,0)$)     {\tiny #4};
		}
	}
}

\tikzset{add2/.style n args={4}{
		minimum width=6mm,
		path picture={
			\draw[black] 
			(path picture bounding box.south) -- (path picture bounding box.north)
			(path picture bounding box.west) -- (path picture bounding box.east);
			\node at ($(path picture bounding box.south)+(0,0.13)$)     {\tiny #1};
			\node at ($(path picture bounding box.west)+(0.13,0)$)      {\tiny #2};
			\node at ($(path picture bounding box.north)+(0,-0.13)$)        {\tiny #3};
			\node at ($(path picture bounding box.east)+(-0.13,0)$)     {\tiny #4};
		}
	}
}

\begin{document}

\title{Low-Complexity Iterative Receiver for \\ Orthogonal Chirp Division Multiplexing }
\author{
		\IEEEauthorblockN{Roberto Bomfin$^\dagger$, Marwa Chafii$^\star$ and Gerhard Fettweis$^\ddagger$, \IEEEmembership{Fellow,~IEEE}} \\
		\IEEEauthorblockA{Vodafone Chair Mobile Communication Systems, Technische Universit\"{a}t Dresden, Germany}$^\dagger$$^\ddagger$ \\ 
		\IEEEauthorblockA{ ETIS, UMR 8051,
Universit\'{e} Paris-Seine, Universit\'{e} Cergy-Pontoise, ENSEA, CNRS, France}$^\star$\\
		\IEEEauthorblockA{roberto.bomfin@ifn.et.tu-dresden.de$^\dagger$, marwa.chafii@ensea.fr$^\star$ and gerhard.fettweis@tu-dresden.de$^\ddagger$} 
		}
\maketitle

\begin{abstract}
This paper proposes a low-complexity iterative receiver for the recently proposed Orthogonal Chirp Division Multiplexing (OCDM) modulation scheme, where we consider a system under frequency-selective channels and constrained to channel state information availability only at the receiver. It has been shown that under these assumptions, OCDM becomes an optimal waveform in terms of performance, i.e., frame error rate (FER), when employing a receiver capable of achieving perfect feedback equalizer (PFE) performance. Thus, this work targets proposing such a receiver for OCDM with low-complexity. Our approach is based on the well accepted minimum mean squared error with parallel interference cancellation (MMSE-PIC), where we derive an approximated equalizer whose complexity is reduced to two fast Fourier transforms (FFTs) per iteration. The FER results reveal that  i) the proposed low-complexity receiver perform as good as the original MMSE-PIC, ii) OCDM performs very closely to PFE, and iii) OCDM has approximately 2.5 dB improvement over OFDM.
%
\end{abstract}

\section{Introduction}\label{sec:introduction}
Robustness is always a concern in the development of new wireless communication technologies. 
For instance, applications such as vehicular communications depend on a reliable physical layer (PHY) in order to deliver services that achieve the required safety constraints \cite{Festag}. Another example are the cellular systems, in which the improvement of the PHY performance allows better quality-of-service and support of new applications.
Therefore, the investigation of novel waveforms is a key aspect for the improvement of current state-of-the-art \ac{OFDM}-based systems such as \ac{LTE} and WiFi. 
In general, OFDM is used in the standards due to its simplicity.
However, the authors in \cite{Bomfin} showed that OFDM is suboptimal under \acp{FSC} when \ac{CSI} is available at the receiver only. 
In contrast, the waveforms that equally spread the symbols in frequency domain provide optimal performance, if non-linear receiver capable of achieving the performance of \ac{PFE} is employed. In other words, the non-linear receiver should be able to resolve the \ac{ISI} caused by the channel, which has been demonstrated in ~\cite{Matthe,Zhang,Ritcey}. 

In \cite{Zhao}, the authors propose the \ac{OCDM} scheme whose symbols are equally spread in frequency. As expected from \cite{Bomfin}, the results of \cite{Zhao} report improvement of OCDM over OFDM in the coded system when a non-linear decision feedback equalizer is used \cite{Benvenuto}, which is a suboptimal receiver \cite{Yilmaz}. 
Therefore, in this paper we aim at complementing the work done in \cite{Bomfin} and \cite{Zhao} by proposing a low-complexity non-linear receiver for \ac{OCDM}. 
We choose the well known \ac{MMSE-PIC} receiver since its performance has been shown to be satisfactory in \ac{ISI} scenarios ~\cite{Studer,Matthe}, which is the case for \ac{OCDM} under \acp{FSC}. In short, the concept of \ac{MMSE-PIC} consists in \acp{LLR} exchange between the equalizer and the \ac{SISO} decoder. 
In order to achieve low-complexity, we design an approximated equalizer that consumes only 2 \acp{FFT} per iteration, which is achieved by assuming that the a-priori symbols are equally reliable. 
Additionally, our receiver also works with the single-carrier waveform and is equivalent to the receiver proposed in \cite{Falconer} for this particular case, since the authors used the MMSE criterion for deriving the forward and backward filters, and they also assumed equal reliability of a-priori symbols.  
In addition, we highlight that although we consider single-input single-out system, we demonstrate that our receiver perfectly works with space time coding (STC) schemes such as Alamouti and delay diversity ~\cite{Nabar,Alamouti,Winters}. 

Our simulations consider an LTE-based system under \ac{ETU} \ac{FSC} channel, assuming CSI at the receiver only. The results show that our low-complexity receiver has almost no performance loss compared to the original MMSE-PIC in terms of \ac{FER}. 
Therefore, we provide experimental evidende that the equal reliability assumption of a-priori symbols used in \cite{Falconer} is indeed accurate.
More importantly, it is also shown that the proposed receiver has practically almost performance as the \ac{PFE}, meaning that OCDM can be considered to be optimal waveform in terms of FER \cite{Bomfin}, since the FER curves related to PFE can be seen as lower bound curves. 
Additionally, we show that OCDM has approximately 2.5 dB gain over OFDM, which clearly shows its the potential. Furthermore, assuming \ac{CRC} after each iteration, we show that for a \ac{FER} less than $10^{-3}$, the amount of necessary iterations is surprisingly low. For instance, for less than 5$\%$ of time the receiver needs to perform more than 1 iteration. This result clearly indicates that the average complexity and processing time of the proposed receiver does not deviate significantly from OFDM, which requires only 1 FFT per frame.

{\it Notation:} unless otherwise stated, vectors are defined with lowercase bold symbols $\mathbf{x}$ whose $i$th element is $(\mathbf{x})_i$.
Matrices are written as uppercase bold symbols $\mathbf{X}$ whose element in $i$th column and $j$th row is $(\mathbf{\mathbf{X}})_{i,j}$. The special matrix $\mathbf{F}$ and vector $\mathbf{1}$ stand for normalized Fourier matrix the column vector of ones, respectively. The size of vector and matrices should be understood from context.
${\rm Tr}\left(\mathbf{X}\right)$ and ${\rm diag}\left(\mathbf{X}\right)$ returns the trace and diagonal elements of $\mathbf{X}$, respectively. $\mathbf{X}^{\rm T}$ and $\mathbf{X}\H$ returns the trace and complex conjugate of $\mathbf{X}$, respectively. Finally, $\mathbb{E}\left\{\cdot \right\}$ stands for the expectation operator.

\section{System Model}\label{sec:system_model}
\subsection{Block Fading Model}
Consider the block fading system
\begin{equation}\label{eq:y}
	\mathbf{y} = \mathbf{H}\mathbf{A}'\mathbf{d} + \mathbf{w},
\end{equation}
where $\mathbf{y} \in \mathbb{C}^N$ is the received signal with $N$ samples, $\mathbf{H} \in \mathbb{C}^{N\times N}$ is the convolution channel matrix, i.e., we assume i) that the channel impulse response remains constant during a block period, and ii) \ac{CP} insertion at transmitter and removal at receiver. $\mathbf{A}' \in \mathbb{C}^{N \times N}$ is the transmitter matrix in time domain and $\mathbf{d} \in \mathcal{S}^{N}$ is the coded data vector with $\mathbb{E}\left\{ \mathbf{d} \mathbf{d}\H \right\} = \mathbf{I}$, and whose elements are taken from the \ac{QAM} constellation set $\mathcal{S}$ with cardinality $|\mathcal{S}| = J$. Finally, $\mathbf{w} \sim \mathcal{CN} (0,\mathbf{I} \sigma^2)$ is the noise component whose variance is defined as $\sigma^2$.

For convenience, consider the received signal of \eqref{eq:y} in frequency domain as
\begin{align}\label{eq:Y}
	\mathbf{Y} & = \mathbf{F}\mathbf{y} \nonumber \\
	& = \mathbf{F} \mathbf{H} \mathbf{F}\H \mathbf{F} \mathbf{A}'\mathbf{d} + \mathbf{F}\mathbf{w} \nonumber \\
	& = \mathbf{\Lambda} \mathbf{A}\mathbf{d} + \mathbf{W},
\end{align}
where $\mathbf{\Lambda} = \mathbf{F} \mathbf{H} \mathbf{F}\H$ is a diagonal matrix whose elements correspond to the channel response in frequency domain. $\mathbf{A} = \mathbf{F} \mathbf{A}'$ is the transmitter matrix in frequency domain and $\mathbf{W} \sim \mathcal{CN} (0,\mathbf{I} \sigma^2)$ is the noise vector in frequency domain.


\subsection{MMSE-PIC receiver}\label{subsec:mmse_pic}
\begin{figure}
	\tikzstyle{block} = [draw, fill=white, rectangle, 
    minimum height=3em, minimum width=6em]
\tikzstyle{multiplier} = [draw,circle,fill=blue!20,add={}{}{}{}] {} 
\tikzstyle{sum} = [draw,circle,fill=blue!20,add2={}{}{}{}] {} 
\tikzstyle{input} = [coordinate]
\tikzstyle{output} = [coordinate]
\tikzstyle{pinstyle} = [pin edge={to-,thin,black}]

\def\windup{
	\tikz[remember picture,overlay]{
		\draw (-0.8,0) -- (0.8,0);
		\draw (-0.8,-0.4)--(-0.4,-0.4) -- (0.4,0.4) --(0.8,0.4);
}}

\centering
\begin{tikzpicture}[auto, node distance=2cm,>=latex']
\node [input, name=input] at (0,0) {};

\node [block, right of=input, fill=blue!20] (equalizer) {Equalizer};
\node [block, right of=equalizer, node distance=3.1cm] (extrinsic_LLR) {Ext. LLRs};
\node [block, below of=extrinsic_LLR, node distance=2cm] (SISO_dec) {SISO Dec.};

\node [output, name=output, right of=SISO_dec,node distance=2cm] {};

\node [block, below of=equalizer, node distance=2cm] (a_priori_calc) {A-priori Stat.};

\node [below of=a_priori_calc,node distance=0.8cm] () {\small \it step 1};
\node [above of=equalizer,node distance=0.8cm] () {\small \it step 2};
\node [above of=extrinsic_LLR,node distance=0.8cm] () {\small \it step 3};
\node [below of=SISO_dec,node distance=0.8cm] () {\small \it step 4};
%
%
%
%

\draw [->] (input) -- node [xshift=-0.42cm]{$(\mathbf{Y}, \mathbf{\Lambda}, \sigma^2)$} (equalizer);
\draw [->] (equalizer.15) -- node[name=bk,yshift=-0.1cm] {$\boldsymbol{\mu}_{\mathbf{d}}^{\rm p}$} (extrinsic_LLR.165);
\draw [->] (equalizer.-15) -- node[name=bp,yshift=-0.1cm] {$\boldsymbol{\Sigma}_{\mathbf{d}}^{\rm p}$} (extrinsic_LLR.-165);%

\draw [->] (extrinsic_LLR) -- node [xshift=0cm]{$L_{s,b}^{\rm e}$} (SISO_dec);

\draw [->] (SISO_dec) -- node [xshift=0cm]{$\hat{b}$} (output);
\draw [->] (SISO_dec) -- node [yshift=0.6cm] {$L_{s,b}^{\rm a}$} (a_priori_calc);
%

\draw [->] (a_priori_calc.55) -- node [xshift=0.1cm] {$\boldsymbol{\Sigma}_{\mathbf{d}}^{\rm a}$} (equalizer.305);
\draw [->] (a_priori_calc.125) -- node [xshift=0.1cm] {$\boldsymbol{\mu}_{\mathbf{d}}^{\rm a}$} (equalizer.235);


%

\end{tikzpicture} 
	\caption{Block diagram of the iterative receiver. Each block represents one step of Subsection \ref{subsec:mmse_pic}.}
	\label{fig:receiver}
\end{figure}
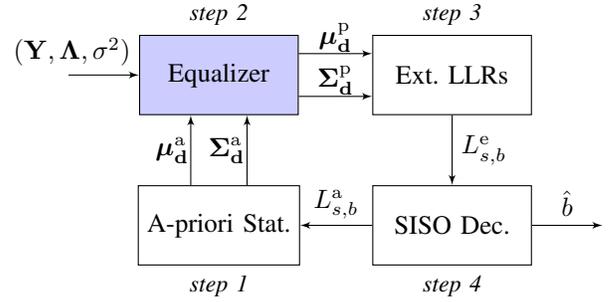

We describe receiver based on the MMSE-PIC with four steps in the following. For completeness, the receiver's block diagram is depicted in Figure \ref{fig:receiver}.
\subsubsection{A-priori statistics of the data vector}
In this step, we compute the a-priori mean and variance of $\mathbf{d}$ based on the a-priori \acp{LLR}, which are provided by the \ac{SISO} decoder. 
Let $L_{s,b}^{\rm a}$ be the a-priori LLR of the $s$th data symbol at the $b$th bit position. 
Then, for the $b$th bit of the $s$th symbol $x_{s,b} \in \left\{0,1\right\}$, its probabilities of assuming 1 or 0 are calculated as ${\rm Pr}\left\{x_{s,b} = 1\right\} = \frac{1}{1+\exp \left(L_{s,b}^{\rm a} \right)}$ and ${\rm Pr}\left\{x_{s,b} = 0\right\} =  \frac{1}{1+\exp \left(-L_{s,b}^{\rm a} \right)}$, respectively. Thus, we can compute the probability mass function of the $s$th data symbol, $d_s \in \mathcal{S}$, as ${\rm Pr}\left\{ d_s = d \right\} = \prod_{b}{\rm Pr}\left\{ x_{s,b} = \mathcal{X}_b(d)^{-1} \right\}$, where $\mathcal{X}_b(d)^{-1}$ is the QAM-to-bit mapping and provides the $b$th bit of the constellation symbol $d$.
Finally, the desired a-priori mean and variance of $\mathbf{d}$ are respectively computed as \cite{Studer}
\begin{align}\label{eq:mu_a}
	(\boldsymbol{\mu}_{\mathbf{d}}^{\rm a})_s & = \sum_{d \in \mathcal{S}}{\rm Pr}\left\{ d_s = d \right\}d, 
\end{align} 
and
\begin{align}\label{eq:Sigma_a}
\left(\boldsymbol{\Sigma}_{\mathbf{d}}^{\rm a}\right)_{s,s} = \sum_{d \in \mathcal{S}} {\rm Pr}\left\{ d_s = d \right\} \left | d - (\boldsymbol{\mu}_{\mathbf{d}}^{\rm a})_s \right |^2.
\end{align}

Notice that in the first iteration, there is no a-priori information available on the LLRs, which means that $L_{s,b}^{\rm a} = 0$ for all $s$ and $b$ due to equiprobability of the transmitted bits. Applying $L_{s,b}^{\rm a} = 0$ in \eqref{eq:mu_a} and \eqref{eq:Sigma_a} leads to $(\boldsymbol{\mu}_{\mathbf{d}}^{\rm a})_s = 0$ and $\left(\boldsymbol{\Sigma}_{\mathbf{d}}^{\rm a}\right)_{s,s} = 1$ for all $s$. 

Finally, the mean quantities are disposed in the vector $\boldsymbol{\mu}_{\mathbf{d}}^{\rm a} = \left((\boldsymbol{\mu}_{\mathbf{d}}^{\rm a})_0, (\boldsymbol{\mu}_{\mathbf{d}}^{\rm a})_1, \cdots, (\boldsymbol{\mu}_{\mathbf{d}}^{\rm a})_{N-1}\right)^{\rm T}$, and the variance quantities are organized in the covariance matrix $\boldsymbol{\Sigma}_{\mathbf{d}}^{\rm a}$ where ${\rm diag}(\boldsymbol{\Sigma}_{\mathbf{d}}^{\rm a}) = (\left(\boldsymbol{\Sigma}_{\mathbf{d}}^{\rm a}\right)_{0,0}, \left(\boldsymbol{\Sigma}_{\mathbf{d}}^{\rm a}\right)_{1,1}, \cdots, \left(\boldsymbol{\Sigma}_{\mathbf{d}}^{\rm a}\right)_{N-1,N-1})$ and the off-diagonal elements are zeros.

\subsubsection{CWCU LMMSE equalization}\label{subsubsec:equalizer}
This step refines the estimate of $\mathbf{d}$ by computing its a-posteriori estimate and error variance. Assuming the model in \eqref{eq:Y}, the \ac{CWCU} LMMSE equalization \cite[eq. (11)]{Matthe} is written as
\begin{equation}\label{eq:mu_p}
\boldsymbol{\mu}_{\mathbf{d}}^{\rm p} \! = \!\boldsymbol{\mu}_{\mathbf{d}}^{\rm a} \!+\! \frac{\mathbf{A}^{\rm H}\mathbf{\Lambda}^{\rm H}\left( \mathbf{\Lambda}\mathbf{A}\mathbf{\Sigma_d^{\rm a}}\mathbf{A}^{\rm H}\mathbf{\Lambda}^{\rm H} \!+\! \sigma^2\mathbf{I}\right )^{-1}\left(\mathbf{Y} \!- \!\mathbf{\Lambda}\mathbf{A}\boldsymbol{\mu}_{\mathbf{d}}^{\rm a} \right )}{{\rm diag \left( \mathbf{A}^{\rm H}\mathbf{\Lambda}^{\rm H}\left( \mathbf{\Lambda}\mathbf{A}\mathbf{\Sigma_d^{\rm a}}\mathbf{A}^{\rm H}\mathbf{\Lambda}^{\rm H} \!+\! \sigma^2\mathbf{I}\right )^{-1}\mathbf{\Lambda}\mathbf{A}\right )}},
\end{equation}
where the division in \eqref{eq:mu_p} is performed component wise in order to ensure that $\boldsymbol{\mu}_{\mathbf{d}}^{\rm p}$ is unbiased, which is achieved by forcing the diagonal elements of the overall equalization matrix to be unitary.

The error variance of $\boldsymbol{\mu}_{\mathbf{d}}^{\rm p}$ \cite[eq. (12)]{Matthe} for the model in \eqref{eq:Y} is given by\footnote{$\mathbf{\Sigma_d^{\rm p}}$ correspond to the diagonal elements of the a-posteriori error matrix. Since $\mathbf{\Sigma_d^{\rm p}}$ is sufficient \cite{Matthe}, we keep it a vector for simplicity.}
\begin{align}\label{eq:Sigma_p}
\boldsymbol{\Sigma}_{\mathbf{d}}^{\rm p} \!  & = \nonumber\\ &\!\frac{\mathbf{1}}{{\rm diag \! \left( \mathbf{A}^{\rm H}\mathbf{\Lambda}^{\rm H}\!\left( \mathbf{\Lambda}\mathbf{A}\mathbf{\Sigma_d^{\rm a}}\mathbf{A}^{\rm H}\mathbf{\Lambda}^{\rm H} \!+\! \sigma^2\mathbf{I}\right )^{-1}\!\mathbf{\Lambda}\mathbf{A}\right )}} - {\rm diag} \! \left(\mathbf{\Sigma_d^{\rm a}} \right ).
\end{align}

Additionally, notice that this step requires the knowledge of channel response in frequency domain as well as the noise power which are $\mathbf{\Lambda}$ and $\sigma^2$, respectively.
\subsubsection{Extrinsic LLR computation}
Based on the outputs of the equalizer, $\boldsymbol{\mu}_{\mathbf{d}}^{\rm p}$ and $\boldsymbol{\Sigma}_{\mathbf{d}}^{\rm p}$, we are able to compute the extrinsic LLRs $L_{s,b}^{\rm e}$ that will serve as input for the SISO decoder. It is given by
\begin{equation}\label{eq:L_e}
L_{s,b}^{\rm e} = \frac{1}{(\boldsymbol{\Sigma}_{\mathbf{d}}^{\rm p})_s}\left( \min_{d \in \mathcal{S}_b^{(0)}}|d-(\boldsymbol{\mu}_{\mathbf{d}}^{\rm p})_s|^2 - \min_{d \in \mathcal{S}_b^{(1)}}|d-(\boldsymbol{\mu}_{\mathbf{d}}^{\rm p})_s|^2 \right ),
\end{equation}
where $\mathcal{S}_b^{(0)}$ and $\mathcal{S}_b^{(1)}$ represents the sets of constellation symbols in which the $b$th bit is 0 or 1, respectively.

It its worth mentioning that equation \eqref{eq:L_e} is an approximation that assumes independence of the noise component in each symbol. Moreover, equation \eqref{eq:L_e} neglects the a-priori knowledge of each bit for simplicity, however only marginal impact on the performance was observed when compared with the optimal LLR computation \cite{Studer}, favoring \eqref{eq:L_e} for implementation.
\subsubsection{SISO Decoder}
This step is responsible for calculating the a-priori LLR of the coded bits, i.e. $L_{s,b}^{\rm a}$, based on the extrinsic LLRs, $L_{s,b}^{\rm e}$. In this work, we use recursive systematic convolutional code with BCJR log-MAP decoder, since the performance of this coding scheme was reported to be satisfactory \cite{Matthe}. In addition, we assume that the interleaving of $L_{s,b}^{\rm a}$ and de-interleaving of $L_{s,b}^{\rm e}$ are done inside the decoder block for simplicity.

When the maximum number of iterations is achieved, the information bits are estimated as $\hat{b}$ by comparing the uncoded LLRs with zero. 

\section{Low-Complexity Equalizer (LCE) for OCDM}\label{sec:low_comp}
In steps 1-4 described in Subsection \ref{subsec:mmse_pic}, the equalization is the most critical step. In particular, the matrix inversion $\left( \mathbf{\Lambda}\mathbf{A}\mathbf{\Sigma_d^{\rm a}}\mathbf{A}^{\rm H}\mathbf{\Lambda}^{\rm H} \!+\! \sigma^2\mathbf{I}\right )^{-1}$ and the element wise division vector ${\rm diag \! \left( \mathbf{A}^{\rm H}\mathbf{\Lambda}^{\rm H}\!\left( \mathbf{\Lambda}\mathbf{A}\mathbf{\Sigma_d^{\rm a}}\mathbf{A}^{\rm H}\mathbf{\Lambda}^{\rm H} \!+\! \sigma^2\mathbf{I}\right )^{-1}\!\mathbf{\Lambda}\mathbf{A}\right )}$ in equations \eqref{eq:mu_p} and \eqref{eq:Sigma_p} have complexity of order $O(N^3)$, which are prohibitive for practical implementation. 


We are interested in providing a low-complexity equalizer for \ac{OCDM}, whose transmitter matrix in frequency domain is given by $\mathbf{A}=\mathbf{\Gamma\H F}$, where $\mathbf{\Gamma}$ is a diagonal matrix with elements $\left( \exp \left(-j \pi n^2/N\right) \right)_{n=0, 1, \cdots, N-1}$ \cite{Zhao}.
In the following, we approach this task by providing approximated and simplified alternatives for the the aforementioned matrix inversion and the element wise division. 
\subsection{Solution for $\left( \mathbf{\Lambda}\mathbf{A}\mathbf{\Sigma_d^{\rm a}}\mathbf{A}^{\rm H}\mathbf{\Lambda}^{\rm H} \!+\! \sigma^2\mathbf{I}\right )^{-1}$}\label{subsec:inverse}
The solution we propose in this work is to replace the a-priori covariance matrix $\mathbf{\Sigma_d^{\rm a}}$ by the scalar 
\begin{equation}\label{eq:sigma_a}
\bar{\sigma}_{\rm a}^2 =\frac{{\rm Tr}(\boldsymbol{\Sigma}_{\mathbf{d}}^{\rm a})}{N},
\end{equation}
which reduces the matrix inversion to $\left( \bar{\sigma}_{\rm a}^2\mathbf{\Lambda}\mathbf{\Lambda}^{\rm H} + \sigma^2\mathbf{I}\right )^{-1}$, since $\mathbf{A}\mathbf{A}\H = \mathbf{I}$ for OCDM. Notice that now we only need to invert a diagonal matrix, which is directly obtained by element wise inversion.

By using $\bar{\sigma}_{\rm a}^2$, we essentially consider that all symbols are equally reliable as done in \cite{Falconer}, which is true only for the first iteration. In other words, we are neglecting the reliability information of individual symbols.
However, in our case of interest, i.e., OCDM under \acp{FSC}, the symbols have the same channel gain since they spread equally in frequency \cite{Bomfin}, leading to similar reliability. Therefore, we can expect that this approximation would cause no significant impact on the performance. The empirical results presented in Section \ref{sec:numetical_evaluations} demonstrate that this assumption holds.

\subsection{Solution for ${\rm diag \! \left( \mathbf{A}^{\rm H}\mathbf{\Lambda}^{\rm H}\!\left( \mathbf{\Lambda}\mathbf{A}\mathbf{\Sigma_d^{\rm a}}\mathbf{A}^{\rm H}\mathbf{\Lambda}^{\rm H} \!+\! \sigma^2\mathbf{I}\right )^{-1}\!\mathbf{\Lambda}\mathbf{A}\right )}$}\label{subsec:normalization}
Again, by replacing $\mathbf{\Sigma_d^{\rm a}}$ by $\bar{\sigma}_{\rm a}^2$ given in \eqref{eq:sigma_a}, we observe that $\mathbf{\Lambda}_{\rm eq} = \mathbf{\Lambda}^{\rm H}\!\left(  \bar{\sigma}_{\rm a}^2\mathbf{\Lambda}\mathbf{\Lambda}^{\rm H} + \sigma^2\mathbf{I}\right )^{-1}\!\mathbf{\Lambda}$ is a diagonal matrix.
In addition, since $\mathbf{A} = \mathbf{\Gamma\H}\mathbf{F}$ with $\mathbf{\Gamma}\H\mathbf{\Gamma}=\mathbf{I}$, we can write the division term as ${\rm diag \left( \mathbf{F}^{\rm H} \mathbf{\Lambda}_{\rm eq} \mathbf{F}\right )}$.
It is easy to see that
\begin{align}
(\mathbf{F}^{\rm H} \mathbf{\Lambda}_{\rm eq} \mathbf{F})_{1,1}  & = \frac{1}{N}\mathbf{1}^{\rm T}\mathbf{\Lambda}_{\rm eq}\mathbf{1} \nonumber \\  & = \frac{{\rm Tr} \left(\mathbf{\Lambda}_{\rm eq}\right)}{N},
\end{align}
since the first row of $\mathbf{F}\H$ and the first column of $\mathbf{F}$ are $\frac{1}{\sqrt{N}}\mathbf{1}^{\rm T}$ and $\frac{1}{\sqrt{N}}\mathbf{1}$, respectively. Moreover, $\mathbf{F}^{\rm H} \mathbf{\Lambda}_{\rm eq} \mathbf{F}$ is circulant, which means that its diagonal has equal elements. Thus, we can write the normalization factor with the scalar
\begin{equation}\label{eq:diag}
\lambda_{\rm norm} = \frac{{\rm Tr} \left(\mathbf{\Lambda}_{\rm eq}\right)}{N}.
\end{equation}
Finally, we can simply replace the denominator of the rightmost term in \eqref{eq:mu_p} with $\lambda_{\rm norm}$.
\subsection{Final equalizer for OCDM}
At this point, we just need to merge the solutions of Subsections \ref{subsec:inverse} and \ref{subsec:normalization} to rewrite equations \eqref{eq:mu_p} and \eqref{eq:Sigma_p}. The results are given by
\begin{equation}\label{eq:mu_p2}
				\boldsymbol{\mu}_{\mathbf{d}}^{\rm p} = \boldsymbol{\mu}_{\mathbf{d}}^{\rm a} + \frac{\mathbf{F}^{\rm H}\mathbf{\Gamma}\mathbf{\Lambda}^{\rm H}\left(\mathbf{Y} - \mathbf{\Lambda}\mathbf{\Gamma}\H\mathbf{F}\boldsymbol{\mu}_{\mathbf{d}}^{\rm p} \right )}{\lambda_{\rm norm}\left( \bar{\sigma}_{\rm a}^2\mathbf{\Lambda}\mathbf{\Lambda}^{\rm H} + \sigma^2\mathbf{I}\right )}
\end{equation}
and
\begin{equation}\label{eq:Sigma_p2}
\mathbf{\Sigma_d^{\rm p}} = \frac{\mathbf{1}}{\lambda_{\rm norm}} - \bar{\sigma}_{\rm a}^2\mathbf{1}.
\end{equation}
It is worth noticing that the simplified equalizer also works for the DFT-Precoded OFDM system, which is a block transmitted single-carrier. In this case, $\mathbf{\Gamma} = \mathbf{I}$.

\subsection{Complexity Analysis}
Since the matrix inversion is done element wise, the computational effort will be governed by the amount of \acp{FFT} operations. In the first iteration, 2 FFTs are necessary, one to compute $\mathbf{Y} = \mathbf{Fy}$ and the left-most inverse FFT. Remember that in the first iteration, $\boldsymbol{\mu}_{\mathbf{d}}^{\rm a} = \mathbf{0}$, thus $\mathbf{F}\boldsymbol{\mu}_{\mathbf{d}}^{\rm a}$ does not count. From the second iteration on, it is also needed 2 FFTs, one for $\mathbf{F}\boldsymbol{\mu}_{\mathbf{d}}^{\rm a}$ and the other for the left-most inverse. Finally, complexity of the low-complexity equalizer can be approximated by $2\times I$ FFTs, where $I$ is the number of iterations. In the next section, we show that the necessary amount of iterations is relatively small.

\subsection{Extension to MIMO-STC}\label{subsec:STC}
In this part, we briefly discuss about the extension of the proposed system to Alamouti and cyclic delay diversity (CDD) STC MIMO \cite{Alamouti,Winters}. Basically, we can design the STC of both schemes such that equivalent channel becomes a regular single-input single-output channel ~\cite{Nabar,Odair}. It means that it can be factored as a diagonal matrix in frequency domain. This is sufficient to model the received signal as \eqref{eq:Y}. The applicability of the proposed receiver is then straightforward.

\section{Numerical Evaluation and Discussion}\label{sec:numetical_evaluations}
\begin{table}[t]
	\centering
	\caption{Simulation Parameters}
	\vspace{-3mm}
	\begin{tabular}{c|c}\toprule
		Parameter & Value  \\ \midrule 
		Available Subcarriers & 1536 \\
		Sampling Rate & 23.04 MHz \\
		Allocated Subcarriers & $12 \times 24 = 288$ \\				
		Bandwidth & 4.32 MHz \\
		OCDM/OFDM Symb./Block & 7 \\
		Channel Model & 3GPP ETU (invariant over a block)   \\ 
		Modulation and Coding & QPSK and 16-QAM with 1/2 Code Rate\\
		Encoder & $\left\{133,171\right\}_8$ Recursive Systematic Conv.  \\
		SISO Decoder & BCJR log-MAP 
		\\ 
		\bottomrule
	\end{tabular}
	\label{tab:parameters}
\end{table}

In this section, we target comparing the performance of the low-complexity equalizer (LCE) with i) the original equalizer, ii) the \ac{PFE}, and iii) with OFDM. For that, we consider a bit-interleaved coded modulation system based on \ac{LTE} with parameters defined in Table \ref{tab:parameters}, where we assume a perfect synchronization and channel knowledge for simplicity. We consider a system with 24 resource blocks of 12 subcarriers each, resulting in 288 subcarriers per block. In addition, the coded bits are interleaved and spread over 7 blocks of 288 subcarriers\footnote{Although our model was not formally designed for more than one transmitted block, one should notice that this configuration is straightforwardly implemented by properly appending the equalizer's inputs and outputs.}. 

\begin{figure}[h!]
	\includegraphics[scale=0.99]{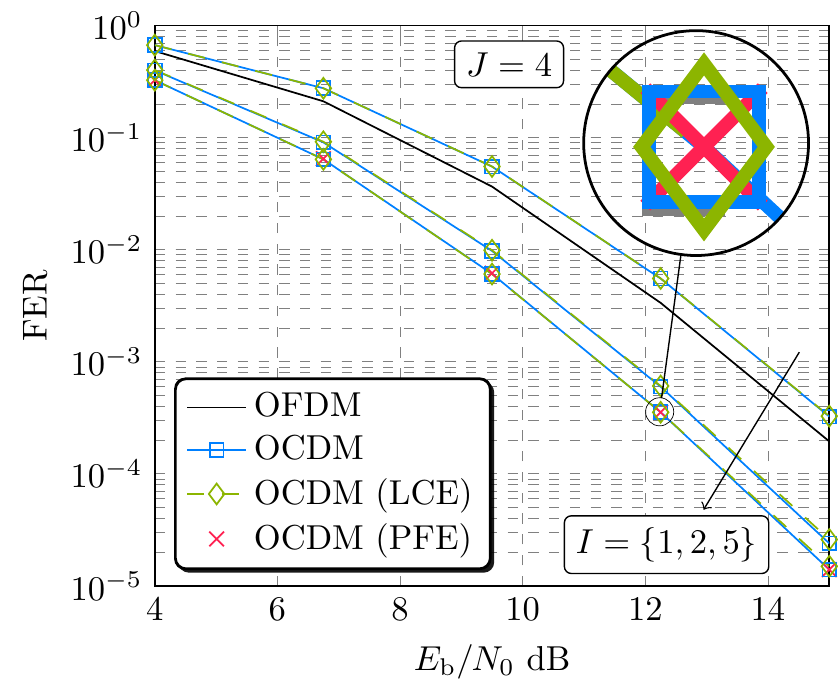}	
	\includegraphics[scale=0.99]{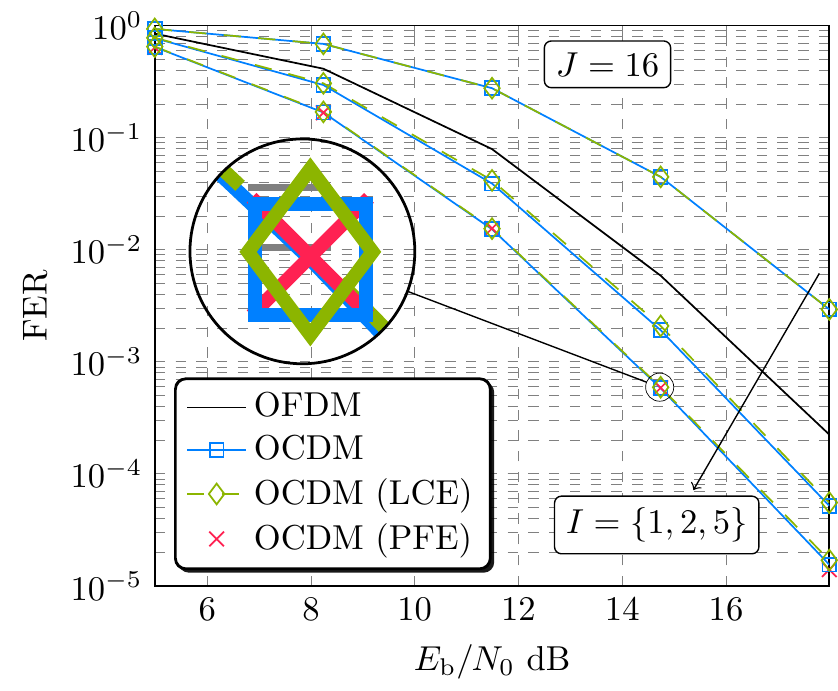}	
	\caption{\ac{FER} for OFDM and OCDM for original equalizer (eqs. \eqref{eq:mu_p} and \eqref{eq:Sigma_p}), low-complexity equalizer (LCE) (eqs. \eqref{eq:mu_p2} and \eqref{eq:Sigma_p2}), and perfect-feedback equalizer (PFE) for $I=\left\{1,2,5\right\}$.}
	\label{fig:FER}
\end{figure}

Figure \ref{fig:FER} shows the \ac{FER} for OFDM and OCDM using the original, LCE and PFE. 
The first thing to notice is that the LCE for OCDM has almost as good performance as the original one, proving that the assumption of no significant performance loss in the derivation of \eqref{eq:mu_p2} is correct. 
Furthermore, one can observe the performance improvement with the number of iterations, showing that the extra signal processing used by the OCDM system is able to correct errors that OFDM can not.
Additionally, for 5 iterations, the performance of OCDM achieves the PFE. Thus, our results are complementary to the theoretical analysis done in \cite{Bomfin}, where the employment of a non-linear receiver that achieves PFE was assumed. 
It means that OCDM can indeed be considered to provide optimal performance. 
Yet, another interpretation of this outcome is that OCDM provides the performance of a system under flat-fading channel, since all symbols under PFE have the same SNR.
Moreover, the performance gain of OCDM over OFDM is around 2.5 dB, which is very significant and clearly demonstrates the potential of OCDM, specially because our proposed receiver can be straightforwardly used with MIMO space-time coding.

%
%
\begin{figure}[h!]
	\includegraphics[scale=0.99]{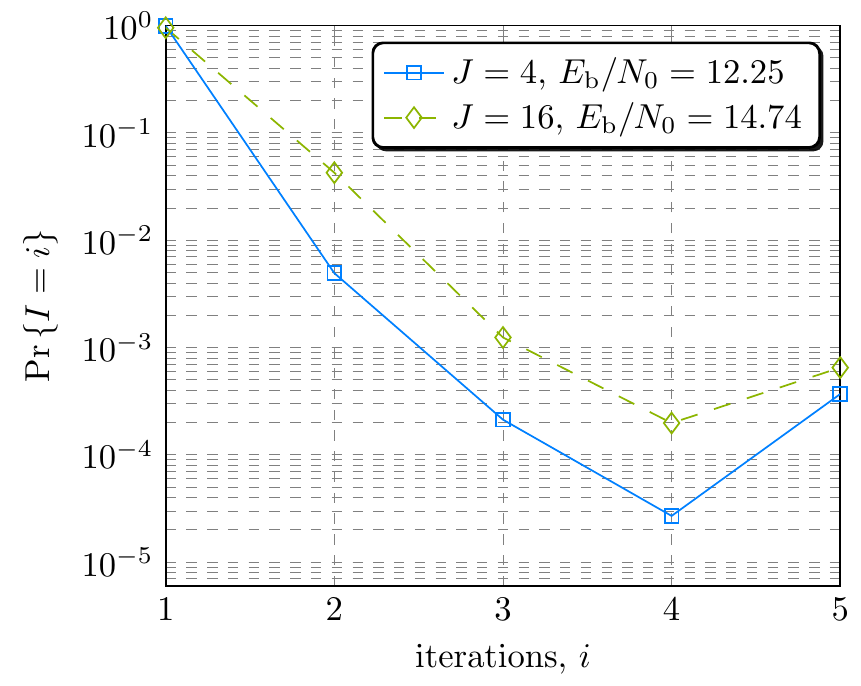}	
	\caption{Histogram of $I$ for $J=4$ with $E_{\rm b}/N_0 = 12.25$ and $J=16$ with $E_{\rm b}/N_0 = 14.74$.}
	\label{fig:HIST}
\end{figure}


In practice, the receiver can perform \ac{CRC} at the end of each iteration, such that the processing can be interrupted as soon as the packet is detected to be correct. Additionally, since the equalizer complexity is proportional to the amount of iterations, i.e., $2 \times I$ FFTs, it is meaningful to analyze the iterations statistically. Thus, Figure \ref{fig:HIST} shows the histogram of iterations when the proposed equalizer is utilized, for $E_{\rm b}/N_0$ equal to $12.25$ dB and $14.74$ dB for $J=4$ and $J=16$, respectively. Notice that these $E_{\rm b}/N_0$ values guarantee a \ac{FER} less than $10^{-3}$ in Figure \ref{fig:FER}. The maximum number of iterations is set to 5, thus if CRC is wrong after the 5th iteration, the packet is declared to be wrong. The results reveal that the receiver detects the packet correctly most of the time in the first iteration. In fact, the probability that the receiver performs 2 or more iterations is less than 5$\%$, showing that the complexity of the proposed system is relatively close to OFDM which consumes only $1\times $FFT. Moreover, one can observe that $I=5$ is more frequent than $I=4$, this happens because when the package is not successfully detected, the receiver necessarily performs 5 iterations. This observation suggests that increasing the maximum number of iterations too much might lead to inefficient hardware usage. 

There are still relevant aspects to be considered. First, we assumed perfect channel estimation and static channel over all 7 transmitted blocks. In fact, in reality these assumptions do not hold. Also, we did not take into account windowing that is used to suppress out-of-band emissions. It means that more realistic conditions should be addressed in future work. Nevertheless, we conjecture that the gap between OCDM and OFDM should increase in the more realistic scenarios, since the iterative receiver can resolve partially the extra interference, whereas for OFDM it would correspond to extra noise. 

In addition, as shown in \cite{Zhao}, OCDM has the same performance as DFT-p OFDM, i.e., single-carrier waveform, when CP covers all channel taps. However, it is also shown in \cite{Zhao} that OCDM outperforms DFT-p OFDM if CP is less than the channel spread for uncoded case. In fact, this concept can also be explored for the coded system with iterative receiver, where the extra interference terms should be considered. We expect that OCDM will also outperform DFT-p OFDM in this case. 

Finally, a systematic analysis of the proposed receiver with STC-MIMO schemes discussed in Subsection \ref{subsec:STC} should also be properly addressed.
\section{Conclusion}
This paper presented a low-complexity approximated MMSE-PIC receiver for OCDM modulation scheme.
The numerical evaluation showed that the proposed receiver has almost no performance loss compared to the general MMSE-PIC, but with significant complexity reduction. 
In addition, our analysis reveals that OCDM can be seen as an optimal waveform under frequency-selective channels with channel state information only available at the receiver. This conclusion is drawn by the fact that our low-complexity receiver achieves the performance of the perfect-feedback equalizer, which is a general lower bound. 
In comparison to OFDM, OCDM presented approximately 2.5 dB gain, which undoubtedly demonstrates the potencial of this waveform for upcoming technologies. 
Furthermore, we commented that our receiver also works for space time coding multiple-input multiple-output schemes, where the diagonalization of the equivalent channel in frequency domain is sufficient for this purpose.

As future research, we propose an analaysis of OCDM under realistic scenarios, i.e., imperfect channel estimation and windowing. In this case, we conjectured that the performance gap between OCDM and OFDM should increase, given that the iterative receiver can resolve extra interference, while for OFDM it would be extra noise.

\section*{Acknowledgments}
The computations were performed at the Center for Information
Services and High Performance Computing (ZIH) at Technische Universit\"{a}t Dresden.

This work was supported by the European Union's Horizon 2020 under grant agreements no. 732174 (ORCA project) and no. 777137 (5GRANGE project).

\bibliography{references}{}
\balance
\bibliographystyle{ieeetr}

\end{document}